\documentclass[preprint]{aastex}

\def\ni{\noindent}

\def\rad{{\rm\,rad}}

\def\yr{{\rm\,yr}}

\def\pomega{\tilde{\omega}}

\begin{document}

\shortauthors{Chiang, Tabachnik, and Tremaine}
\shorttitle{Apse Alignment in $\upsilon$ And}

\title{Apsidal Alignment in Upsilon Andromedae}

\author{E.~I.~Chiang\altaffilmark{1,2}, S.~Tabachnik\altaffilmark{3}, \&
S.~Tremaine\altaffilmark{3}}

\altaffiltext{1}{Hubble Fellow}
\altaffiltext{2}{Institute for Advanced Study, School of Natural
Sciences, Einstein Drive, Princeton, NJ~08540, USA}
\altaffiltext{3}{Princeton University Observatory, Peyton Hall,
Princeton, NJ~08544-1001, USA}

\email{chiang@ias.edu, serge@astro.princeton.edu, tremaine@astro.princeton.edu}

\begin{abstract}
One of the parameters fitted by Doppler radial velocity measurements
of extrasolar planetary systems is $\omega$, the argument of
pericenter of a given planet's orbit referenced
to the plane of the sky. Curiously, the $\omega$'s
of the outer two planets orbiting Upsilon Andromedae
are presently nearly identical:
$\Delta \omega \equiv \omega_D - \omega_C = 4\fdg8 \pm 4\fdg8 \, (1\sigma)$.
This observation is least surprising if planets C and D
occupy orbits that are seen close to edge-on
($\sin i_C$, $\sin i_D \gtrsim 0.5$) and whose mutual inclination
$\Theta$ does not exceed $20\degr$.
In this case, planets C and D inhabit a secular resonance
in which $\Delta \omega$ librates about $0\degr$ with
an amplitude of $\sim$$30\degr$ and a period
of $\sim$$4\times 10^3 \yr$. The resonant configuration spends
about one-third of its time with $|\Delta\omega| \leq 10\degr$.
If $\Theta \gtrsim 40\degr$,
either $\Delta \omega$ circulates or the system is unstable.
This instability is driven by the Kozai mechanism which couples
the eccentricity of planet C to $\Theta$ to
drive the former quantity to values approaching unity.
Our expectation that $\Theta \lesssim 20\degr$
suggests that planets C and D formed
in a flattened, circumstellar disk, and
may be tested by upcoming astrometric measurements with
the FAME satellite.
\end{abstract}


\section{INTRODUCTION}
\label{intro}

Upsilon Andromedae ($\upsilon$ And) is a Sun-like star harboring at least three
planetary
companions (Butler et al.~1999). The star's radial velocity variations
are fitted approximately by the superposition of three Keplerian
sinusoids, each of which takes the form

\begin{equation}
V = K [\cos (f + \omega) + e \cos \omega] \, .
\label{dopp}
\end{equation}

\ni Here

\begin{equation}
K = \frac{m \sin i}{M_{\ast} + m} \sqrt{\frac{G (M_{\ast} + m)}{a (1-e^2)}} \;
,
\label{doppk}
\end{equation}

\ni $f$ is the true anomaly of a given planet, $i$ is the unknown
inclination between this planet's orbit plane and the plane of the sky,
and the quantities $G$, $M_{\ast} = 1.3 M_{\odot}$,
$m$, $a$, and $e$ take their usual meanings.
Definitions and current fitted values of orbital parameters are listed in
Table \ref{fit}, as kindly supplied by G. Marcy and D. Fischer (2001).

\placetable{fit}

The quantity $\omega$ in equation ($\ref{dopp}$) and Table \ref{fit}
is a given
planet's argument of pericenter referenced to the plane
of the sky:

\begin{equation}
\omega = {\rm sign} [ (\mathbf{\hat{n}} \times \mathbf{\hat{e}}) \cdot
\mathbf{\hat{l}} ] \arccos (\mathbf{\hat{n}} \cdot \mathbf{\hat{e}}) \in
(-\pi,\pi] \label{omegasky}
\end{equation}

\ni where

\begin{equation}
\mathbf{\hat{n}} \equiv \frac{\mathbf{\hat{s}} \times \mathbf{\hat{l}}}{\sqrt{1
- (\mathbf{\hat{s}} \cdot \mathbf{\hat{l}})^2}} \; . \label{hatn}
\end{equation}

\ni Let the origin be positioned at the barycenter
of $\upsilon$ And and take the sky plane to contain
this origin. Then $\mathbf{\hat{s}}$ is the unit
vector pointing from the origin to the observer; $\mathbf{\hat{l}}$
is the unit vector parallel to the planet's orbital angular momentum;
$\mathbf{\hat{n}}$ is the unit vector directed from the origin
to the node of the orbit on the sky plane at which the
planet is approaching the observer; and $\mathbf{\hat{e}}$ is the
unit vector pointing from the origin
to the pericenter of the planet's orbit. Figure \ref{defgeo} illustrates
the geometry. A single value of $\omega$
corresponds to a volume of allowed parameter space
swept out by vectors $\mathbf{\hat{s}}$, $\mathbf{\hat{l}}$, and
$\mathbf{\hat{e}}$.

\placefigure{defgeo}
\begin{figure}
\plotone{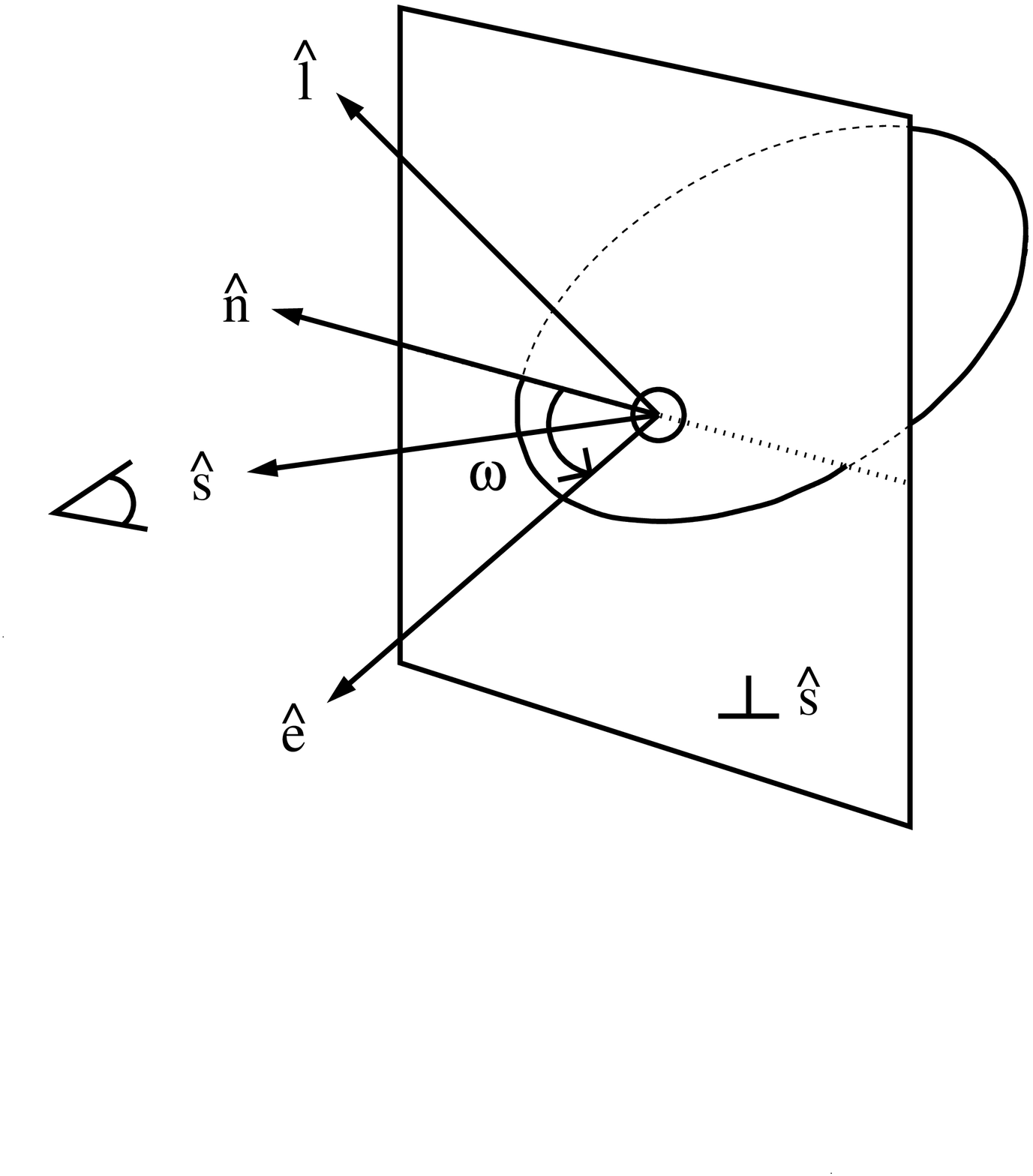}
\end{figure}

Curiously, the $\omega$'s of the outer two planets C and D
are presently nearly
identical: $\Delta \omega = \omega_D - \omega_C = 4\fdg8 \pm 4\fdg8 \,
(1\sigma)$. Let us define
$\Theta = \arccos\,(\mathbf{\hat{l}}_C \cdot \mathbf{\hat{l}}_D)$
to be the mutual inclination between the two orbit planes
of planets C and D.
If we assume for the moment that $\Theta = 0\degr$, so
that the orbits are co-planar and ``co-rotating,'' then
the observation that $|\Delta \omega| \ll 1 \rad$ implies
the near perfect alignment of orbital pericenters. We may ask
whether the alignment is merely coincidental,
or whether a dynamical mechanism exists to lock the
apsidal lines together.

Precedents for apsidal locking exist in the Solar System.
Eccentric planetary rings precess rigidly about Uranus and Saturn;
the inner and outer elliptical edges of a given ring maintain
the same line of apsides by a balance of forces due to the
quadrupole field of the central planet, ring self-gravity,
and interparticle collisions (Chiang \& Goldreich 2000). Of
greater relevance to the case of $\upsilon$ And is the
observation that the periapsides of some asteroids
librate (undergo small oscillations) about the apsidal
line of Jupiter (Milani \& Nobili 1984). Such
asteroids inhabit a secular apsidal resonance
in which their time-averaged apsidal precession
rates match that of Jupiter.

Planets C and D of $\upsilon$ And might occupy a similar
apsidal resonance. How might we use this
possibility to constrain the origin of these planets?
Suppose that such a resonance
operates if $\Theta = 0\degr$, and suppose that it does
not operate otherwise. In the latter
case, suppose that $\Delta \omega$ circulates (cyclically
runs the gamut between $0\degr$ and $360\degr$),
so that the probability of finding $|\Delta \omega| \ll 1\rad$
at any given time is small.
Given these assumptions, the observation today
that $|\Delta \omega| \ll 1\rad$ might lead us to suspect that
$\Theta = 0\degr$.
Co-planarity would argue, in turn, that planets C and D
formed from a flattened, circumstellar disk.

Even if the resonant dynamics are fully understood,
this line of reasoning is suggestive
at best.
A quantitative, Bayesian assessment of the {\it a posteriori}
probability distribution for $\Theta$ given
the observation that $|\Delta \omega| \ll 1\rad$
requires knowledge of the {\it a priori}
probability distribution for $\Theta$.
The latter distribution depends upon the
unknown conditions of the formation of these
two planets---the very object of our inferences.

This paper examines the long-term
evolution of the orbits of planets D, C,
and to a limited extent B,
within the multi-dimensional parameter space of
present-day viewing and orbital geometries
that are permitted by the Doppler observations.
We seek to identify regions
in this parameter space---in particular, all ranges of the
mutual inclination $\Theta$---that correspond to
orbital evolutions for which $|\Delta \omega| \ll 1\rad$
for all time; we suggest that the planets of $\upsilon$ And
reside in regions that are characterized by such apsidal resonances.
The volume of parameter space to be surveyed is substantial;
for instance, if $\Theta \neq 0\degr$, then
$\omega_{D} = \omega_{C}$ does not even necessarily imply that
$\mathbf{\hat{e}}_C$, $\mathbf{\hat{e}}_D$, $\mathbf{\hat{l}}_C$,
and $\mathbf{\hat{l}}_D$ lie in the same plane.
Equality of $\omega$'s merely reflects the equality
of angles which may well be referenced to different nodes
and which may therefore have no direct physical relation.

Previous explorations of this parameter space by Rivera \&
Lissauer (2000ab) and Stepinski, Malhotra, \& Black (2000)
have indeed uncovered the existence of a secular apsidal
resonance between planets C and D in the $\Theta = 0\degr$
case.
Furthermore, considerations of dynamical stability
presented by these authors limit values of $\sin i$ in the
$\Theta = 0\degr$ case to be greater than $\sim$0.3.
Stability considerations also argue against $\Theta \gtrsim 60\degr$
(Stepinski et al.~2000).

Our analysis extends and improves upon these previous results
in a number of respects. First, we take explicit account of the dependence
of $\omega$ on viewing geometry.
To our knowledge, the only study to mention this
dependence is that of Rivera \& Lissauer (2000a),
who incorporate it in their calculations of co-planar,
counter-rotating orbits ($\Theta = 180\degr$).
Second, we systematically explore
all of the parameter space spanned by possible
orbital configurations
of planets C and D and by possible lines-of-sight;
previous studies explore only a fraction of this space.
Third, we focus on the evolution
of $\Delta \omega$ as a means of distinguishing
between likely and unlikely orbital configurations.
As we shall see, the existence of a secular
apsidal resonance characterized by small libration
amplitude in select regions of parameter space
suggests tighter constraints on $\Theta$
than considerations of stability alone.
Finally, the Doppler-fitted values of orbital parameters that we
employ are improved over those used in previous
studies due to the increased number of data points
($N = 189$ radial velocity measurements as of January 2001;
Marcy \& Fischer 2001).

In \S\ref{cop}, we take $\Theta$
to be zero and study the secular evolution of the orbits of
planets C and D while neglecting the presence of the innermost
planet B.
We recover analytically the secular resonance found numerically
by Rivera \& Lissauer (2000ab) and Stepinski et al.~(2000) and
describe its physical character. In \S\ref{mut},
we allow $\Theta \neq 0\degr$ while continuing to ignore planet B.
All viewing/orbital configurations for which $\Delta\omega$
potentially librates about $0\degr$ are identified.
In \S\ref{planb}, we ask which of the resonant
configurations found in \S\ref{mut} are likely to remain
resonant with the inclusion of planet B.
Section \ref{conc} marshals our results
to argue that the planets of $\upsilon$ And most likely
originated in a circumstellar disk.

\section{Planets C and D: $\Theta = 0\degr$}
\label{cop}

In the co-planar, co-rotating case, the individual $\omega$'s
represent longitudes of pericenters referenced to a common nodal
vector $\mathbf{\hat{n}}$ and a common orbital plane.
We ask in this case whether $\Delta \omega$ circulates
or librates,
and investigate how the answer depends on uncertainties
in the orbital parameters, including $\sin i$.

\subsection{Analytic Description}
\label{at}

Long-term variations in the
eccentricities and apsidal longitudes of the outer
two planets are qualitatively well
described by the classical
secular solution of Laplace-Lagrange (L-L; see, e.g., Murray \& Dermott 1999).
Let $q = m_D / m_C = 2.03 \pm 0.05\;(1\sigma)$, $\alpha = a_C / a_D = 0.3238
\pm 0.0009\;(1\sigma)$, and let $\beta = b^{(2)}_{3/2} (\alpha) / b^{(1)}_{3/2}
(\alpha) = 0.476 / 1.19 = 0.399 \pm 0.001\;(1\sigma)$ be
the ratio of Laplace coefficients (Brouwer \& Clemence 1961). These
quantities are evaluated using the orbital parameters of
Marcy \& Fischer (2001). Then

\begin{eqnarray}
\label{ll}
e_C \exp i\omega_C & = & e_{C+} \exp i(g_+ t + \gamma_+) \, + \, e_{C-} \exp
i(g_- t + \gamma_-) \\
e_D \exp i\omega_D & = & e_{D+} \exp i(g_+ t + \gamma_+) \, + \, e_{D-} \exp
i(g_- t + \gamma_-) \label{ll4}
\end{eqnarray}

\ni where the frequencies $g_+$ and $g_-$, and ratios
$e_{D+} / e_{C+}$ and $e_{D-} / e_{C-}$,
are constants specified by the masses and secularly invariant
semi-major axes of the two planets:

\begin{eqnarray}
g_{\pm} = \frac{\pi m_C}{4 M_{\ast} P_C} \alpha^2 b^{(1)}_{3/2}(\alpha) \left\{
q + \sqrt{\alpha} \pm \sqrt{ (q - \sqrt{\alpha})^2 + 4q\sqrt{\alpha}\beta^2}
\right\} \approx \left\{ \begin{array}{l}
2\pi / (7000 \sin i \yr)\\
2\pi / (30000 \sin i \yr)
\end{array}
\right. \label{freq} \\
\left( \frac{e_D}{e_C} \right)_{\pm} = \frac{q - \sqrt{\alpha} \mp \sqrt{ (q -
\sqrt{\alpha})^2 + 4q\sqrt{\alpha}\beta^2}}{2q\beta} \approx \left\{
\begin{array}{c}
-0.14\\
1.9
\end{array}
\right. \; . \label{rat}
\end{eqnarray}

\ni Here $P_C$ is the orbital period of planet C.
These equations neglect terms of order $e^3$ and $(m/M_*)^2$.
The Laplace-Lagrange
description is valid for small values of the mutual inclination
$\Theta$ provided $\omega$ is replaced by $\pomega$, the
longitude of pericenter referenced to the invariable plane
(the usual ``dog-leg'' angle defined in, e.g., Murray \& Dermott 1999).

In writing equations (\ref{ll})--(\ref{rat}), we have omitted
terms due to the third, innermost planet B. Provided planet B
remains on an orbit that is
more than a factor of 10 smaller than the orbits
of the outer two bodies, its time-averaged potential acts mainly
as a static quadrupole moment. This quadrupole could add corrections
up to order $(m_B/m_C)(a_B/a_C)^2$ $\sim$ a few percent
to our expressions for the eigenfrequencies
(\ref{freq}) and eigenvector amplitude ratios (\ref{rat}).
In this section and \S\ref{mut},
we neglect the effects of planet B
for clarity of presentation and computational ease.
We incorporate quantitatively the effects of planet B in \S\ref{planb}.

The remaining four constants $\gamma_{\pm}$, $e_{C+}/e_{C-}$, and
$e_{D+}/e_{D-}$ require specification of the initial eccentricities
and apsidal longitudes.
If initial conditions are such that
$|e_{C+} / e_{C-}| \ll 1$ and $|e_{D+} / e_{D-}| \ll 1$,
then equations (\ref{ll})--(\ref{ll4}) imply that
the two planets precess in lockstep at frequency $g_-$
with fixed eccentricities and aligned pericenters.
%
The present-day orbital parameters of $\upsilon$ And are such that
most, but not all, of the power
is in the ``$-$'' eigenmode. Setting
$\omega_C = 0\degr$, $\omega_D = 4\fdg8$, $e_C =0.25$, and
$e_D = 0.34$ at $t=0$, we obtain with a numerical root-finder:

\begin{equation}
\gamma_+ = -10\degr, \;\;\;\;\; \gamma_- = 4\fdg3, \;\;\;\;\; e_{D+}/e_{D-} =
-0.030, \;\;\;\;\; e_{C+}/e_{C-} = 0.41 \; .
\label{finalanswer}
\end{equation}

\ni The evolution described by equations
(\ref{ll})--(\ref{finalanswer}) is plotted in
Figures \ref{ew}abc.
On average, both $\omega_D$ and $\omega_C$ advance at the
same rate $g_-$.
In addition, however, the admixture of power from the
``$+$'' mode ($e_{C+}/e_{C-}$ is not substantially
smaller than unity) implies that $\omega_C$ librates about $\omega_D$
at a faster frequency $\sim$$g_+$. The argument of
pericenter of planet D librates less
since that planet carries the bulk ($\sim$75\%) of the angular momentum
in the system.

\placefigure{ew}\notetoeditor{Please have this Figure as large as possible
in the published version, e.g. full page size.}
\begin{figure}
\plotone{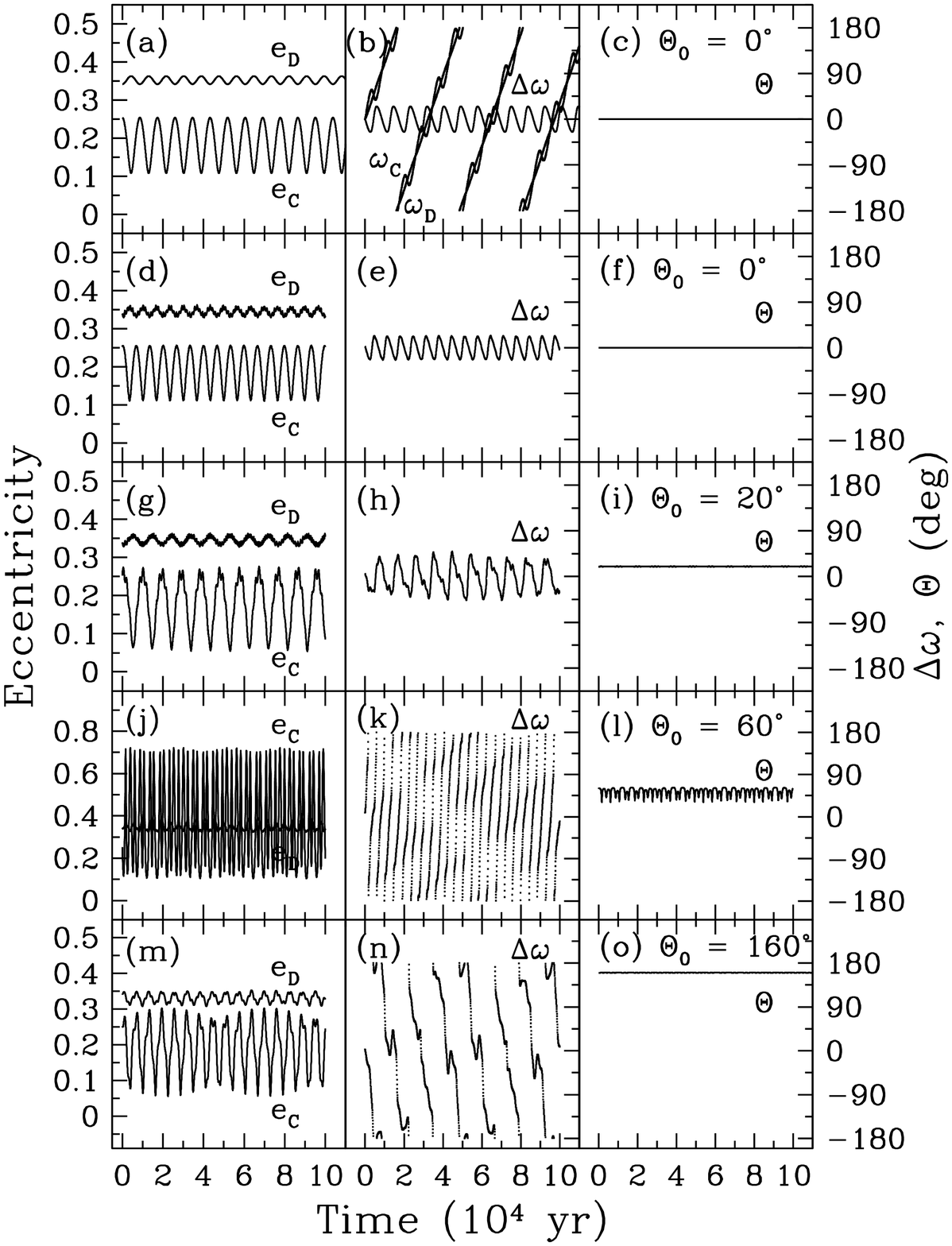}
\end{figure}

The apsidal libration amplitude in L-L theory is a function
of the initial individual eccentricities ($e_{C0}$ and $e_{D0}$),
the initial difference in apsidal longitudes ($\Delta \omega_0$),
$\alpha$, and $q$.
For planets C and D,
most of the uncertainty in the L-L libration amplitude
arises from uncertainties in the fitted eccentricities.
Figure \ref{contour}a exhibits the variation of L-L libration amplitude
with $e_{C0}$ and $e_{D0}$, for
fixed $\alpha = 0.324$, $K_D / K_C = 1.19$,
and $\Delta \omega_0 = 0\degr$.
Allowance is made for the fact that for fixed $\alpha$
and $K_D/K_C$, the observed
$q = m_D / m_C \propto \sqrt{1-e_{D0}^2} \, / \sqrt{1-e_{C0}^2}$
according to equation (\ref{doppk}).
We take $\Delta \omega_0$ to be $0\degr$
because this approximation permits the
immediate solution of equations (\ref{ll})--(\ref{rat}).
Variations in libration amplitude with
$\Delta \omega_0 \in [-4\fdg8, 14\degr]\,(2\sigma {\rm\,confidence\,interval})$
are typically less than 10\% (other parameters being held fixed);
these variations are small compared to
the $\sim$50\% variation in libration amplitude
associated with the $2\sigma$ confidence region
in fitted eccentricities.
Variations in libration amplitude
associated with uncertainties in
$\alpha$ and $K_D / K_C$
are negligibly small, at the level
of $\sim$0.3\% and $\sim$3\%, respectively.
Since the L-L libration amplitude
is sensitive to the mass ratio but not to the individual masses,
it is independent of $\sin i$ in the co-planar case.

\placefigure{contour}
\begin{figure}
\plotone{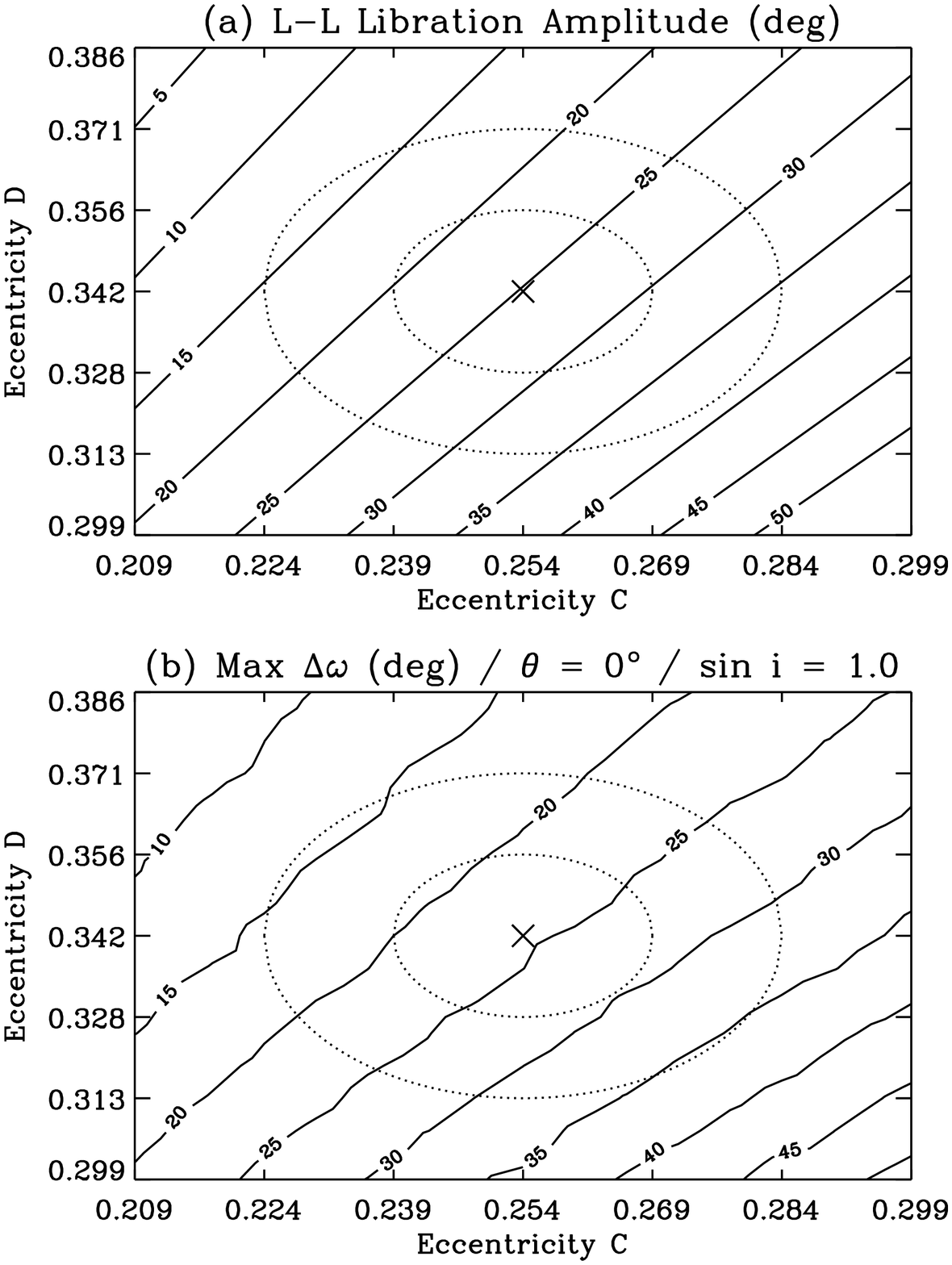}
\end{figure}

The L-L amplitude associated with the
best-fit eccentricities is 25$\degr$.
The corresponding fraction of time that the
system spends with $| \Delta \omega | \leq 10\degr$
equals $P_{time} = 0.26$.
Within the $2\sigma$ error ellipse of fitted
eccentricities, the amplitude ranges
from $14\degr$ ($P_{time} = 0.52$) to
$38\degr$ ($P_{time} = 0.16$

We conclude from this analytic study that if planets C and D occupy
co-planar, co-rotating orbits, then they inhabit a secular resonance
for which $\Delta \omega$ librates about $0\degr$ with
an amplitude of $25\degr \pm 6\degr \; (1\sigma)$.
In other words, if $\Theta = 0\degr$,
the observation today that $|\Delta \omega| \leq 10\degr$
should not be regarded as surprising, since the best-fit
system spends about one-quarter of its time in this state.
These libration parameters
change slightly as we incorporate more realistic
effects in subsequent sections (e.g., the
presence of planet B is included in \S\ref{planb}).

\subsection{Numerical Solution}
\label{ns}
We verify the analytical considerations above by numerically
integrating the orbits of planets C and D using the SWIFT software package
developed by Levison \& Duncan (1994). We employ
their mixed variable symplectic integrator, which is
derived from the algorithm invented by
Wisdom \& Holman (1991). As before, the effects of planet B
are neglected. The timestep is sufficiently small
($\Delta t = 3$ days) that the total energy in
the system is conserved to within 1 part in $10^7$
after $t = 10^6 \yr$.
Initial conditions are taken from
the best-fit parameter values in Table \ref{fit}.
The starting epoch of our
integrations was chosen arbitrarily to be
JD 2450160.1 days. Since the effects in which we
are primarily interested are secular (orbit-averaged), our results
should be qualitatively insensitive to choice of starting epoch.
We have verified that none of our conclusions
changes by choosing an alternative starting epoch
of JD 2451160.1 days.
Moreover, since substantial changes in the
osculating orbital elements of planets C
and D occur only over secular timescales that are long
compared to the duration of the radial velocity
observations, the errors in the fitted orbital
parameters in Table \ref{fit} introduced by the assumption
of fixed Keplerian ellipses are no more than
several percent (Laughlin 2001; see also Laughlin \& Chambers 2001).

As seen in Figures \ref{ew}def, the behavior of
$\Delta \omega$ with time for $\sin i = 1$ computed using SWIFT
is similar to that predicted by L-L.
For smaller values of $\sin i \in [ 1 , 0.3 ]$ and correspondingly
greater planetary masses, the system
remains in the secular apsidal resonance, but short
period terms in the disturbing potentials effectively increase
the amplitude of libration.
Thus, the fraction of time for which $|\Delta \omega| \leq 10\degr$
decreases from $P_{time} = 0.30$ to 0.20 as $\sin i$ decreases
from 1 to 0.3.
For $\sin i \lesssim 0.3$,
we find, as do Rivera \& Lissauer (2000b) and Stepinski et al.~(2000),
that the system is unstable.

Figure 3b plots the maximum value of
$\Delta \omega$ over $10^5 \yr$ as a function of the initial
eccentricities for $\sin i = 1$. The contours
are similar to those predicted by L-L,
with deviations introduced by short period
terms and higher order secular terms.
The range in $P_{time}$ within the $2\sigma$ error
ellipse of fitted eccentricities is $[0.20, 0.54]$.

\section{Planets C and D: $\Theta \neq 0\degr$}
\label{mut}

How does the secular evolution of $\Delta \omega$
change with mutual orbital
inclination, $\Theta$?
In particular, for what ranges of $\Theta$ can
planets C and D occupy an apsidal resonance?
We answer these questions using Monte-Carlo
sampling techniques and numerical orbit integrations.
As in \S2, we neglect the effects of planet B.
We begin by delineating the parameter
space spanned by viewing geometries and orbital configurations
of planets C and D that are allowed by the Doppler observations.
Subsequent subsections describe (1) the method by which we sample
the allowed volume of parameter space
and (2) the dynamics that play out
within this volume.

\subsection{Geometrical Considerations}
\label{geocon}

Let the set of ``initial orientations'' refer to the
set of $\{ \mathbf{\hat{l}}_C, \mathbf{\hat{l}}_D, \mathbf{\hat{e}}_C,
\mathbf{\hat{e}}_D, \mathbf{\hat{s}} \}$
such that $\mathbf{\hat{e}}_D \cdot \mathbf{\hat{l}}_D = \mathbf{\hat{e}}_C
\cdot \mathbf{\hat{l}}_C = 0$ and
$\mathbf{\hat{l}}_D \cdot \mathbf{\hat{l}}_C = \cos \Theta_0$.
The subscript ``$0$'' denotes the initial or present-day value;
as we shall see, the mutual inclination varies with
time in most circumstances.
These orbital and viewing orientation vectors sweep out
a volume in the following five dimensions.
Without loss of generality, we erect a set of Cartesian
coordinate axes $\mathbf{\hat{x}} = \mathbf{\hat{e}}_C$,
$\mathbf{\hat{z}} = \mathbf{\hat{l}}_C$, and
$\mathbf{\hat{y}} = \mathbf{\hat{z}} \times \mathbf{\hat{x}}$.
Then the five dimensions of initial orientation space
are $\Theta_0, \Omega_0, \chi_0$---the initial inclination, initial longitude
of ascending node, and initial argument of pericenter, respectively,
of the orbit of planet D upon that of planet C---and $\phi, \delta$---the
azimuthal and polar angles of $\mathbf{\hat{s}}$.
In the absence of observational constraints, the angles
$\Omega_0, \chi_0, \phi \in [0, 2\pi)$, while $\Theta_0, \delta \in [0,\pi]$.

The observed present-day values of $\omega_C$
and $\omega_D$ carve out
a subset in initial orientation space which we call the
set of ``allowed initial orientations.'' We combine
this set with the best-fit values of the remaining
measured orbital parameters of planets C and D
to define the set of ``allowed initial conditions.''
Each member of the set of allowed initial conditions
may be used as initial conditions for the SWIFT orbit integrator. We refer
to the resultant evolution as a ``scenario.''

\subsection{Sampling}
\label{samp}

For each of 19 values of $\Theta_0$ spaced uniformly
between $0\degr$ and $180\degr$, we randomly
generate sets of $(\Omega_0, \chi_0, \phi, \delta)$ using
the following normalized probability distribution functions.
Isotropy of physical space demands that
$\partial P/\partial \phi = 1 / 2\pi$,
$\partial P/\partial (\cos \delta) = 1 / 2$,
and $\partial P/\partial \Omega_0 = 1 / 2\pi$;
i.e., $\phi$, $\cos \delta$, and $\Omega_0$ are uniformly
distributed over their respective domains.
The probability distribution function
for $\chi_0$, like that for $\Theta_0$, is unknown (and may well
depend on other parameters such as $\Omega_0$),
reflecting the unknown physics of the formation
of these planets.
We assume for simplicity
that $\chi_0$ is uniformly distributed over its
domain; i.e., $\partial P / \partial \chi_0 = 1/2\pi$.

For a given $\Theta_0$, those randomly generated values of
$(\Omega_0, \chi_0, \phi, \delta)$ which
satisfy the observed present-day values of
$\omega_C \in [ -111\fdg1 , -106\fdg9 ]\,(1\sigma)$
and $\omega_D \in [ -106\fdg9 , -101\fdg1 ]\,(1\sigma)$
[as computed using equation (\ref{omegasky})]
constitute our sample of allowed initial orientations.
For our choice of $\partial P / \partial \chi_0$,
we find empirically that the fraction of initial orientations
that are allowed does not vary with $\Theta_0$.
At every $\Theta_0$, the first $N = 45$, randomly
generated, allowed initial conditions are employed
as input conditions for the SWIFT orbit integrator.
We believe that our choice for $N$ adequately
samples parameter space since we have taken
$N = 135$ at several
values of $\Theta_0$ and find none of our conclusions
below to be changed.
As discussed in \S\ref{ns}, the starting epoch
of all integrations was JD 2450160.1 days.

\subsection{Dynamics}
\label{dyn}

Each orbit integration lasts
$t = 10^6 \yr$. In addition to recording
the usual orbital parameters such as the
eccentricities and mutual inclination,
we also compute $\Delta \omega$
according to equation (\ref{omegasky}).
The integration spans $10^2$--$10^3$
libration/circulation periods of $\Delta \omega$;
instabilities usually manifest themselves on shorter
timescales. A system is considered to
be unstable if either
planet collides with the central star
or if either planet is ejected from the
system. In practice, we take the latter criterion
to be satisfied if the semi-major axis
of the orbit exceeds 20 AU; we have checked
{\it a posteriori} that stable scenarios
preserve semi-major axes to within $\sim$10\%.
Integrating $45 \times 19 = 855$ scenarios,
each for $t = 10^6\yr$ with a timestep of
$\Delta t = 3$~days, requires 6 CPU days on a DEC Alpha workstation.

Secular libration of $\Delta \omega$ about $0\degr$ is a typical
outcome for $\Theta_0 \lesssim 30\degr$. Figures \ref{ew}ghi
display the evolution of eccentricities, $\Delta \omega$, and
$\Theta$ in one example scenario for which $\Theta_0 = 20\degr$.
The caption to Figure \ref{ew} lists the employed values
of $\sin i_C$ and $\sin i_D$.
The
gross features of the evolution are the same as those for the co-planar
case (cf. Figures \ref{ew}def).
Because the Doppler-measured $\omega$ is sensitive
to both $\mathbf{\hat{e}}$ and $\mathbf{\hat{l}}$,
$\Delta \omega$ is modulated not just at the single usual
frequency $\sim$$g_+$, but also at the two remaining
secular precession frequencies, $\sim$$g_-$ and the nodal
precession frequency

\begin{equation}
\dot{\Omega} \approx -\frac{\pi m_C}{2M_{\ast}P_C} \alpha^2
b^{(1)}_{3/2}(\alpha) (q + \sqrt{a}) \approx - 2\pi / (6000 \yr) \; .
\end{equation}

\ni The above expression is derived to lowest order in $\Theta$ and
the numerical evaluation takes $\sin i_C \approx \sin i_D \approx 1$
(as is appropriate for the scenario in Figures \ref{ew}ghi).
In Figure \ref{ew}i, the mutual inclination $\Theta$ remains constant
at $20\degr$; secular invariance of $\Theta$ obtains for
2-planet systems with small $\Theta$ and small
eccentricities (Brouwer \& Clemence 1961).

Circulation of $\Delta \omega$, instability,
and, more rarely, libration of $\Delta \omega$ are
three possible outcomes for $40\degr \lesssim \Theta_0 \lesssim 90\degr$.
The qualitatively distinct
character of the disturbing function at
such large inclinations was first explored by Kozai (1962),
who described the secular evolution of highly inclined
orbits of asteroids in Jupiter's gravitational field.
Here planet D, the main repository of angular
momentum, behaves as Jupiter, and planet C behaves
as an asteroid. For mutual inclinations
such that $\cos^2 \Theta \lesssim 3/5 \,(\approx \cos^2 39\fdg2)$,
the system may inhabit the Kozai resonance, for which
$\mathbf{\hat{l}}_C$ and $\mathbf{\hat{e}}_C$
precess about the axis parallel to the total angular momentum
vector of the system at a common frequency $\sim$$|\dot{\Omega}|$, while
$\mathbf{\hat{e}}_D$ precesses
about that axis at the slower frequency $\sim$$g_-$.
The result is that $\Delta \omega$ circulates at $\sim$$|\dot{\Omega}|$.
Figures \ref{ew}jkl show
how the eccentricities, $\Delta \omega$, and $\Theta$
evolve under the Kozai mechanism.
Note how $e_C$ and $\Theta$ undergo larger variations than in
the small $\Theta$ case; $e_C$ and $\Theta$ can be coupled
in the $\cos^2 \Theta \lesssim 3/5$ case such that both vary secularly
at frequency $\sim$2$|\dot{\Omega}|$ while keeping Kozai's integral,
$\sqrt{1-e_C^2} \cos \Theta$, approximately constant.
The secular driving of $e_C$ to values
approaching unity causes 90\%
of our sampled scenarios having $\Theta_0 \approx 90\degr$ to self-destruct
within $t = 10^6\yr$.

For $\Theta_0 \gtrsim 90\degr$, the angular momenta of planets
C and D, projected onto the total angular momentum vector
of the system, point in opposite directions. Consequently, secular
exchanges of angular momenta between the two planets
cause their eccentricities to rise and fall together
(Rivera \& Lissauer 2000a). For
$90\degr \lesssim \Theta_0 \lesssim 140\degr$,
the synchronized oscillations of the eccentricities compounds
the destabilizing effects of coupled
$e_C$ and $\Theta$ to render
the majority of scenarios unstable.
For $150 \degr \lesssim \Theta_0 \lesssim 180 \degr$,
$e_C$ and $\Theta$ decouple
and $\Delta \omega$ circulates stably. A typical example of the
evolution in the nearly co-planar, counter-rotating
case is showcased in Figures \ref{ew}mno.

\placefigure{prodist}
\begin{figure}
\plotone{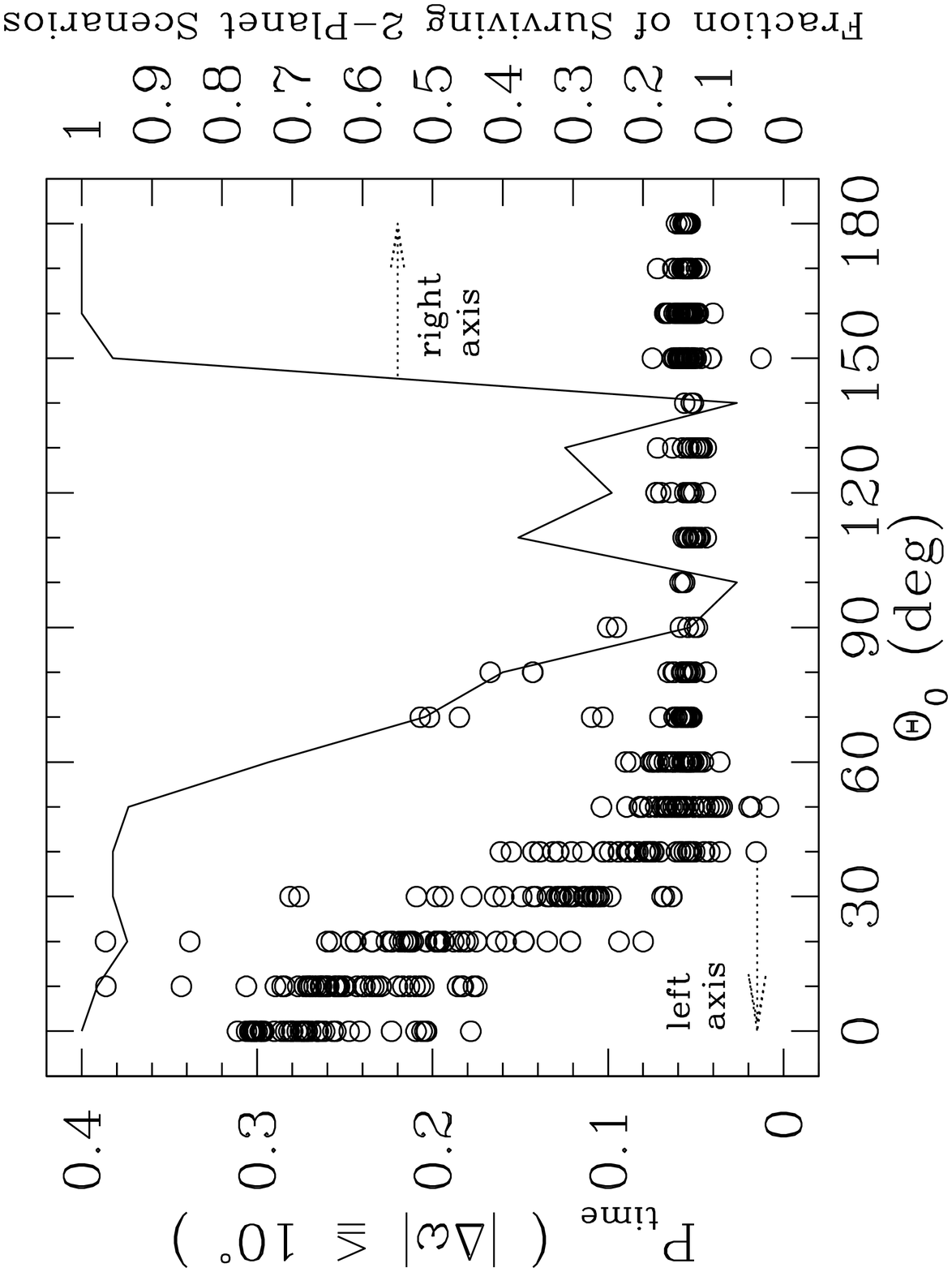}
\end{figure}

Figure \ref{prodist} summarizes the effects we have discussed
so far. The solid line traces the fraction of allowed
initial conditions, sampled according to the assumptions
stated in \S\ref{samp}, that generate 2-planet scenarios which
survive for $t = 10^6 \yr$.
Nearly co-planar scenarios, whether
co-rotating or counter-rotating, betray
little or no sign of instability.
Those co-rotating
systems at $\Theta_0 \leq 30\degr$ that
are unstable all have $\sin i \lesssim 0.3$ for either planet C or D.
Our estimated fraction of surviving systems is an upper limit
because we have neglected thus far the presence of planet B
and because there may be instabilities that require
$t > 10^6 \yr$ to develop.
We expect the survival fraction to be most severely
overestimated at $40\degr \lesssim \Theta_0 \lesssim 140\degr$,
inclinations for which $e_C$ and $\Theta$
undergo large, coupled variations; values of
$e_C \gtrsim 0.93$ can cause planets C and B to undergo destabilizing
close encounters.

Open circles in Figure \ref{prodist} mark
$P_{time}\,(|\Delta \omega| \leq 10\degr)$,
the fraction of time a surviving sampled
scenario spends with $|\Delta \omega| \leq 10\degr$.
For $\Theta_0 \lesssim 20\degr$, the
two planets are locked
in a secular apsidal resonance for
which $P_{time}$ can be as high as
$0.3$--$0.4$. For $\Theta_0 \gtrsim 40\degr$,
$\Delta \omega$ typically circulates so that
$P_{time} = 20\degr / 360\degr = 0.055$.
The specific distribution of values of $P_{time}$
at a given $\Theta_0$ depends upon
our assumed sampling function
for $\chi_0$, the initial
argument of pericenter (see \S\ref{samp}).
For given $\Theta_0$, lower values of $P_{time}$
typically correspond to lower values of $\sin i_C$
and $\sin i_D$ (see \S\ref{ns}).
Maximum values of $P_{time} \approx 0.39$ are attained
at $\Theta_0 = 10$--$20\degr$.

Surprisingly, a secular apsidal resonance
exists at $\Theta_0 \approx 70$--$80\degr$
for which $P_{time} \approx 0.2$.
The evolution within this ``anomalous''
resonance at high $\Theta_0$
is characterized by rapid oscillations
in $\Delta \omega$, $e_C$, $e_D$, and
$\Theta$ at twice the nodal precession
frequency, $\sim$2$|\dot{\Omega}|$.
The eccentricity of planet C is periodically driven
by secular perturbations to values as high as 0.95.
Given such large values of $e_C$,
we suspect that
including the presence of
planet B may render these anomalously resonant configurations
unstable. We confirm this suspicion
in the next section.

\section{Planets B, C, and D}
\label{planb}

In the previous section we found those allowed
initial conditions involving only planets C and D
that result in libration of $\Delta \omega$.
Here we add the effects of the third, innermost
planet B to these resonant configurations,
in the expectation that some of the resultant
3-planet systems will not be stable.

The inclusion of planet B introduces
a number of complications. First,
three extra dimensions are added to the space
of allowed initial orientations: $\Theta_{B0}$,
$\Omega_{B0}$, and $\chi_{B0}$---respectively,
the initial inclination, initial
longitude of ascending node, and initial
argument of pericenter of the orbit of planet B
upon the reference orbit of planet C (see \S\ref{geocon}).
Since the present-day orbit of planet B appears nearly
indistinguishable from a circle (see Table \ref{fit}), we
can take $e_{B0} = 0$ to eliminate $\chi_{B0}$
as a dimension. Even with this simplifying
assumption, the volume of additional parameter space
to be surveyed is, in principle, substantial.
Computational needs are further exacerbated
by the fact that the orbital period of planet
B is 4.6 days; this necessitates a commensurately
short computational timestep. Finally,
planet B is sufficiently close to its
parent star that contributions to the
planet's apsidal and nodal precession
rates due to stellar oblateness
and general relativistic effects are
not negligible compared to contributions
from the outer two planets.

We proceed with a simplified program. First,
we select a set of allowed initial orientations
for planets C and D which give rise to
apsidally resonant, 2-planet scenarios.
At each $\Theta_0$, we select the one
2-planet scenario that generates the maximum value
of $P_{time}\;(|\Delta\omega| \leq 10\degr)$,
provided this quantity exceeds 0.15.
As evident in Figure \ref{prodist}, the values of $\Theta_0$ which
give rise to such scenarios are
$0\degr$, $10\degr$, $20\degr$, $30\degr$, $40\degr$,
$70\degr$, and $80\degr$. Next, to
each member of this set of allowed initial orientations,
we add the following orientation angles for planet B:
$\Theta_{B0} = \Theta_0$ and $\Omega_{B0} = \Omega_0$.
That is, the orbit of planet B is assumed to
be initially circular and
to lie in the orbit plane of the most massive planet, D.
Finally, this augmented set of allowed initial
orientations is combined with the remaining
measured orbital parameters for all three planets
and employed as initial conditions for the SWIFT orbit integrator.
We employ a timestep of $\Delta t = 0.20$ days (5\% of
the orbital period of planet B) and
run each integration for $t = 10^6 \yr$.
The effects of stellar oblateness and
general relativity are ignored.

While the above procedure does not permit
accurate determination of the long-term
orbital evolution of planet B, we believe it
does capture qualitatively
the effect of planet B on the
possibility of resonant apsidal lock
between planets C and D. We expect
the innermost planet to represent
primarily a potential means of destroying
the apsidal resonance, as the secular
growth of $e_C$ takes planet C
disruptively close to planet B.

\placefigure{bcd}\notetoeditor{Please have this Figure
as large as possible in the published version, e.g., full page
size.}
\begin{figure}
\plotone{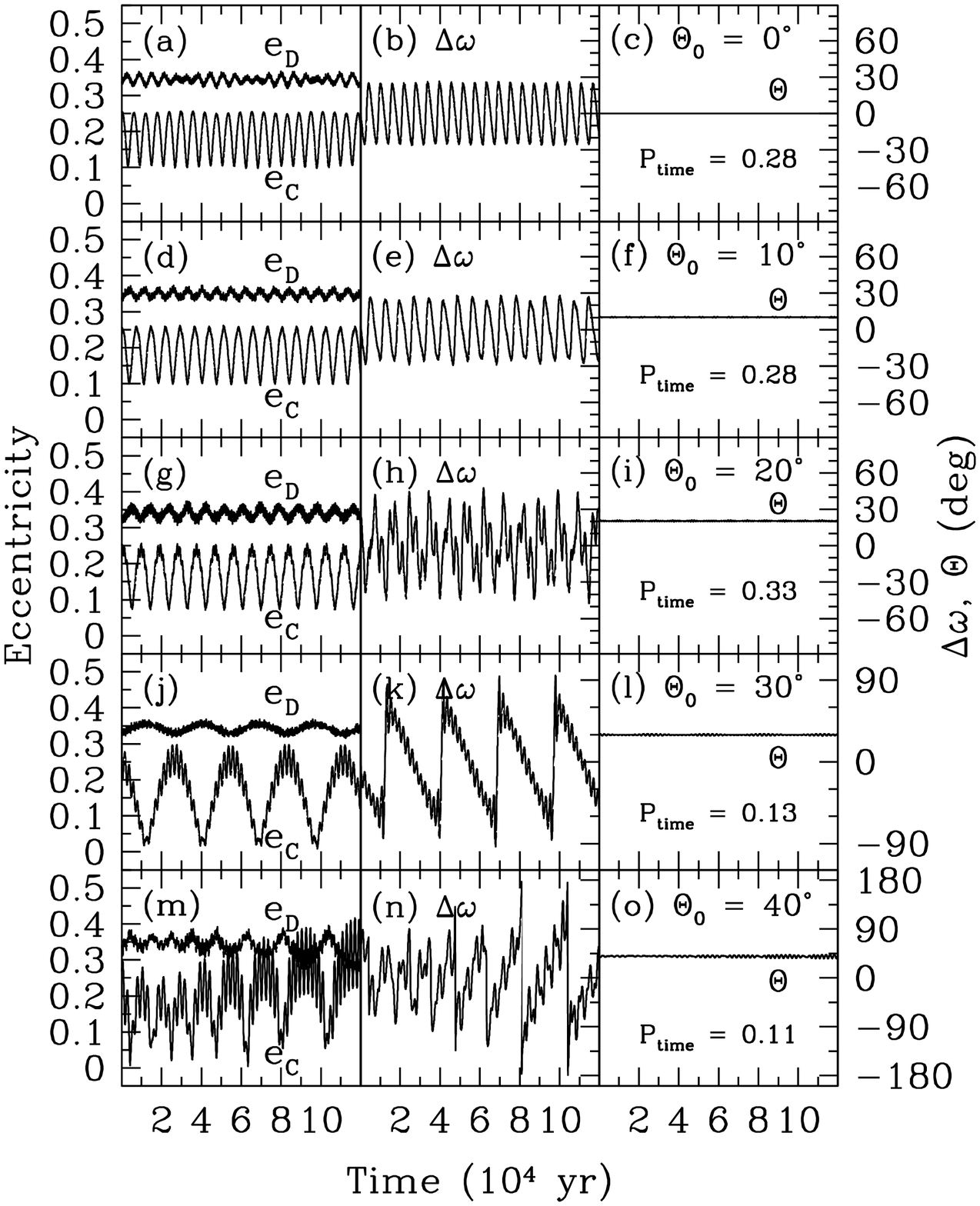}
\end{figure}

Scenarios for which all three
planets survive the entire duration
of the integration are shown in Figure \ref{bcd}.
Only scenarios for which $\Theta_0 = \Theta_{B0} \leq 40\degr$
qualify. For $\Theta_0 \geq 70\degr$,
the eccentricity of planet C is
secularly driven to such high values that
planet B is perturbed
by planet C into a star-crossing orbit
in $t\sim 10^4\yr$.
Thus, the ``anomalous'' resonant
configurations at $\Theta_0 = 70\degr$--$80\degr$
found in \S\ref{dyn} do not survive the
inclusion of planet B.
A similar fate probably befalls
the scenario for which
$\Theta_0 = 40\degr$; an orbit integration
of duration $t > 10^6\yr$ is
required to confirm this suspicion.

For $\Theta_0 \leq 30\degr$,
planets C and D can remain in stable resonant lock
despite the presence of planet B.
Values of $P_{time}\;(|\Delta \omega| \leq 10\degr)$
in 3-planet scenarios
are somewhat reduced from their respective values in 2-planet scenarios.
Values of $P_{time} \approx 0.28$--$0.33$ obtain
for $\Theta_0 \leq 20\degr$; at $\Theta_0 = 30\degr$,
$P_{time}$ drops to 0.13.

\section{Concluding Remarks}
\label{conc}

Our results suggest that the outer two planets of
Upsilon Andromedae occupy
nearly co-planar, co-rotating orbits. The lines of evidence
are as follows:

\begin{enumerate}

\item Finding the Doppler measured $\omega$'s of planets
C and D to be nearly equal
today is least surprising if the mutual orbit inclination
$\Theta_0 \lesssim 20\degr$ (Figures \ref{prodist} and \ref{bcd}).
If $\Theta_0 \lesssim 20\degr$,
planets C and D can be locked in a secular apsidal resonance
for which $\Delta\omega \equiv \omega_D - \omega_C$
librates about $0\degr$ with an amplitude of approximately~$30\degr$.
The corresponding fraction of time that
$|\Delta \omega| \leq 10\degr$ is
$P_{time} \approx 0.30$. These libration parameters are
subject to change with future refinements in
the values of the present-day fitted eccentricities
(Figure \ref{contour}).
By contrast, if
$\Theta_0 \gtrsim 40\degr$, typically $\Delta \omega$ circulates
and $P_{time} \approx 20\degr / 360\degr = 0.055$ for the
relatively few allowed initial conditions that
give rise to stable systems.

\item The secular apsidal resonance at $\Theta_0 \lesssim 20\degr$
limits variations in eccentricities and thereby affords
the system stability (Figure \ref{ew}). By contrast,
if $40\degr \lesssim \Theta_0 \lesssim 90\degr$, secular
driving of the eccentricity of planet C to values
approaching unity via the Kozai resonance fosters
close approaches and concomitant
instability.
If $90\degr \lesssim \Theta_0 \lesssim 140\degr$, the destabilizing
effects of coupled $e_C$ and $\Theta$
are abetted by synchronized driving of the eccentricities.

\end{enumerate}

\ni We note that co-planar, counter-rotating configurations
($\Theta \approx 180\degr$) are allowed by argument 2
but disfavored by argument 1.
Furthermore, values of
$\sin i_C$, $\sin i_D \gtrsim 0.5$ are favored,
not only for reasons of stability, but also because
such values are typical of scenarios for which
the fraction of time spent with small $|\Delta\omega|$
is maximized.

The preference for small mutual inclination, based
on combining considerations of stability with the
likelihood of finding $|\Delta \omega| \ll 1\rad$,
is consistent with
the origin of these planets in a flattened, circumstellar disk.
How does our expectation that
$\Theta_0 \lesssim 20\degr$ compare with orbital
inclinations in related contexts?
The opening angle of hydrostatically flared protoplanetary disks
at stellocentric distances of $\sim$1 AU is
$\sim$6$\degr$ (see, e.g., Chiang \& Goldreich 1997).
The planets in the Solar System, with the understandable exception
of Pluto (Malhotra 1998), occupy orbits whose mutual inclinations
do not exceed about 7 degrees; Mercury aside, mutual
inclinations do not exceed about 3 degrees.

By contrast, scenarios for the formation of extrasolar planets
that do not involve a primordial disk may predict a
substantial population of systems for which $\Theta \gtrsim 150\degr$.
For example, Papaloizou \& Terquem (2000)
have proposed that extrasolar planets might gravitationally
fragment from, and scatter out of, a circumstellar spherical
cloud of gas. Planetary systems having a wide
distribution of initial mutual inclinations
from $\Theta_0 = 0\degr$ to $180\degr$ might be expected to
form. From the stability considerations
elucidated above, we might expect a bimodal population
of systems to survive---those for
which $0\degr \lesssim \Theta \lesssim 40\degr$
and those for which
$150\degr \lesssim \Theta \lesssim 180\degr$---intermediate inclinations
having been eliminated via the destructive Kozai effect.

The final word on the actual values of orbital inclinations
in extrasolar planetary systems will likely come from
combining the Doppler radial
velocity data with astrometric measurements,
or, in rare cases, from transit observations
(Mazeh et al.~2000; Queloz et al.~2000).
The amplitudes of stellar wobbles induced by planets C and D
are ($0.0943 / \sin i_C$) mas and ($0.576 / \sin i_D$) mas,
respectively (Pourbaix 2001). The star $\upsilon$ And
is sufficiently bright ($V = 4.2$) that the upcoming astrometric
satellite FAME (Full-Sky Astrometric Mapping Explorer)
should yield at least a $2\sigma$ detection of the
astrometric signature of planet C.

\acknowledgements
We thank Geoff Marcy and Debra Fischer for generously
providing updated values of the fitted orbital parameters
of $\upsilon$ And, Fathi Namouni for useful conversations,
and Greg Laughlin for a prompt and thoughtful referee's report.
Support for EIC was provided by NASA through a
Hubble Fellowship grant awarded by the Space Telescope
Science Institute, which is operated by the Association
of Universities for Research in Astronomy, Inc., for
NASA under contract NAS 5-26555.
S.~Tabachnik acknowledges
support from an ESA Research Fellowship.
S.~Tremaine acknowledges support
from NASA grants NAG5-7310 and NAG5-10456.

\newpage

\figcaption[chiang.fig1.eps]{Defining geometry for the Doppler-measured
argument of pericenter $\omega$, the angle
between pericenter and the node of the planetary orbit on the plane
of the sky. The sky plane selected is that which
passes through the system barycenter.
The vector $\mathbf{\hat{n}}$ lies in the planes of the
orbit and of the sky. The vector $\mathbf{\hat{e}}$
lies in the orbit plane.
The node selected is the one for which the planet travels towards
the observer. The argument $\omega$ advances along the orbit
plane in the direction of increasing true anomaly.
\label{defgeo}}

\figcaption[chiang.fig2.eps]{Sampling of 2-planet scenarios in which the
orbits
of planets C and D are initially inclined by the value of $\Theta_0$
indicated. Each horizontal row of panels displays the evolution
of $e_C$, $e_D$, $\Delta\omega = \omega_D - \omega_C$, and
$\Theta = \arccos \; (\mathbf{\hat{l}}_C \cdot \mathbf{\hat{l}}_D)$
for a given scenario. For panels (a)--(c), the evolution is
computed using the classical, analytic solution of Laplace-Lagrange
for $\Theta = 0\degr$. For the remaining panels, the evolution
is computed by numerical integration. Libration of $\Delta \omega$
is a typical outcome for $\Theta_0 \lesssim 20\degr$, while
circulation of $\Delta \omega$ and large secular variations
of $e_C$ and $\Theta$ via the Kozai mechanism are possible outcomes
for $40\degr \lesssim \Theta_0 \lesssim 140\degr$.
For $\Theta_0 \gtrsim 150\degr$, $\Delta \omega$ circulates
while $e_C$ and $e_D$ vary modestly and synchronously. Note in panel (f)
how $\Delta \omega$ is modulated at a number of frequencies,
reflecting its dependence on the 4 precessing
vectors $(\mathbf{\hat{l}})_{C,D}$ and $(\mathbf{\hat{e}})_{C,D}$.
For panels (a)--(f), $\sin i  = 1$ and timescales scale
as $\sin i$. For panels (g)--(i), $\sin i_C = 0.99$
and $\sin i_D = 0.93$. For panels (j)--(l), $\sin i_C = 1$
and $\sin i_D = 0.73$. For panels (m)--(o), $\sin i_C = 0.78$
and $\sin i_D = 0.87$.
\label{ew}}

\figcaption[chiang.fig3.eps]{Contours of constant libration
amplitude in degrees as a function of possible present-day
fitted eccentricities, for $\Theta = 0\degr$ and $\sin i = 1$.
The cross indicates the best-fit eccentricities from
Marcy \& Fischer (2001), while
the inner and outer dotted ellipses enclose the
$1\sigma$ and $2\sigma$ confidence regions for
the eccentricities, respectively.
Panel (a) is calculated using Laplace-Lagrange theory
while panel (b) plots the maximum value of $\Delta \omega$
over $t = 10^5\yr$ in numerical integrations using SWIFT.
\label{contour}}

\figcaption[chiang.fig4.eps]{{\it Open circles, left-hand ordinate}: Fraction
of time that surviving, sampled, 2-planet scenarios
spend with $|\Delta \omega| \leq 10\degr$.
The observation today that $|\Delta\omega| \leq 10\degr$
is least surprising if the mutual inclination $\Theta$ between
the orbit planes of planets C and D is less than or equal to $20\degr$.
At these modest inclinations, the two planets inhabit
a secular apsidal resonance in which $\Delta\omega$ librates
about $0\degr$ with an amplitude of $\sim$$25\degr$.
The cluster of points at $\Theta = 70\degr$--$80\degr$
and relatively large $P_{time} \approx 0.20$ reflects
the existence of an ``anomalous'' apsidal 2-planet
resonance. In \S\ref{planb}, we discuss
how the inclusion of the third, innermost
planet B disrupts the anomalous resonance.
{\it Solid line, right-hand ordinate}: Fraction of 2-planet
scenarios that survive the
$10^6 \yr$ duration of the integration. Nearly co-planar
orbits, whether co-rotating or counter-rotating, betray
little sign of instability. Scenarios at intermediate inclinations
tend to be unstable because the Kozai effect at $\Theta \gtrsim 40\degr$,
abetted by synchronous driving of eccentricities
at $\Theta \gtrsim 90\degr$,
drives the eccentricity of planet C to such high values that
planet C collides with the star.
\label{prodist}}

\figcaption[chiang.fig5.eps]{Three-planet scenarios in which planets
B and D initially occupy co-planar orbits while planet C initially
occupies an orbit inclined by the value of $\Theta_0$ indicated.
Each horizontal row of panels displays the evolution of $e_C$, $e_D$,
$\Delta\omega = \omega_D - \omega_C$, and
$\Theta = \arccos \; (\mathbf{\hat{l}}_C \cdot \mathbf{\hat{l}}_D)$
for a given scenario. Scenarios for which $\Theta_0 \leq 20\degr$
preserve apsidal libration of $\Delta\omega$
despite the presence of planet B and spend
the maximum fraction of time with $|\Delta\omega| \leq 10\degr$:
$P_{time} \approx 0.28$--$0.33$. At $\Theta_0 = 30\degr$,
the amplitude of apsidal libration is at its largest,
approximately $90\degr$.
At $\Theta_0 \geq 40\degr$, $\Delta\omega$ eventually circulates.
For panels (a)--(c), $\sin i_B = \sin i_C = \sin i_D = 0.89$.
For panels (d)--(f), $\sin i_B = \sin i_D = 0.99$ and $\sin i_C = 0.95$.
For panels (g)--(i), $\sin i_B = \sin i_D = 0.79$ and $\sin i_C = 0.55$.
For panels (j)--(l), $\sin i_B = \sin i_D = 0.99$ and $\sin i_C = 0.97$.
For panels (m)--(o), $\sin i_B = \sin i_D = 0.90$ and $\sin i_C = 0.44$.
\label{bcd}}

\newpage
\begin{deluxetable}{cccccccc}
\rotate
\tablewidth{0pt}
\tablecaption{Fitted Orbital Parameters of Upsilon Andromedae (Marcy \& Fischer
2001)\tablenotemark{a} \tablenotemark{b}\label{fit}}
\tablehead{
\colhead{Planet} &
\colhead{$P$ (days)}	& \colhead{$T_{peri}$ (JD)} &
\colhead{$e$}	& \colhead{$\omega$} &
\colhead{$K$ (m/s)}	& \colhead{$m \sin i$ ($M_{J}$)} &
\colhead{$a$ (AU)}
}
\startdata
B & 4.61706 ($7\times 10^{-5}$) & 2450001.6 (NA)\tablenotemark{c} & 0.015
(0.009) & 32.1 (NA)\tablenotemark{c} & 71.0 (0.7) & 0.69 & 0.059 \\
C & 241.14 (0.22) & 2450160.1 (1.9) & 0.254 (0.015) & -109.0 (2.9) & 55.6 (0.9)
& 1.96 & 0.828 \\
D & 1309.13 (5.13) & 2450044.0 (11.3) & 0.342 (0.015) & -104.2 (3.8) & 66.2
(1.2) & 3.98 & 2.557 \\
\enddata
\tablenotetext{a}{Based on fitting $N = 189$ radial velocity points, with
rms residuals of 11.5 m/s.}
\tablenotetext{b}{The orbital period is $P$, the Julian date of pericenter
passage is $T_{peri}$, the eccentricity is $e$, the planetary mass
is $m$ measured in Jupiter masses $M_J$, and the semi-major axis is $a$.
Remaining variables are
defined in the text. Uncertainties ($1\sigma$) in fitted values
are enclosed in parentheses.}
\tablenotetext{c}{Uncertainties in $\omega$ and $T_{peri}$ have
little meaning for the nearly circular orbit of planet B.}
\end{deluxetable}

\end{document}